\documentclass[aps,prb,twocolumn,amsmath,reprint,longbibliography,superscriptaddress]{revtex4-2}
\usepackage{graphicx}
\usepackage{bm}
\usepackage{amssymb}
\usepackage{amsmath}
\usepackage{upgreek}

\begin{document}

\title{Evidence of anisotropic three-dimensional weak-localization in TiSe$_{2}$ nanoflakes}

\author{Xiaocui Wang}
\affiliation{Key Laboratory of Advanced Optoelectronic Quantum Architecture and Measurement, Ministry of Education, School of Physics, Beijing Institute of Technology, Beijing 100081, China}
\affiliation{Micronano Center, Beijing Key Lab of Nanophotonics and Ultrafine Optoelectronic Systems, Beijing Institute of Technology, Beijing 100081, China}

\author{Yang Yang}
\affiliation{Center for Joint Quantum Studies and Department of Physics, School of Science, Tianjin University, Tianjin 300350, China}
\affiliation{Tianjin Key Laboratory of Low Dimensional Materials Physics and Preparing Technology, Department of Physics, Tianjin University, Tianjin 300350, China}

\author{Yongkai Li}
\affiliation{Key Laboratory of Advanced Optoelectronic Quantum Architecture and Measurement, Ministry of Education, School of Physics, Beijing Institute of Technology, Beijing 100081, China}
\affiliation{Micronano Center, Beijing Key Lab of Nanophotonics and Ultrafine Optoelectronic Systems, Beijing Institute of Technology, Beijing 100081, China}

\author{Guangtong Liu}
\affiliation{Beijing National Laboratory for Condensed Matter Physics, Institute of Physics, Chinese Academy of Sciences, Beijing 100190, China}
\affiliation{School of Physical Sciences, University of Chinese Academy of Sciences, Beijing 100049, China}
\affiliation{Songshan Lake Materials Laboratory, Dongguan, Guangdong 523808, China}

\author{Junxi Duan}
\affiliation{Key Laboratory of Advanced Optoelectronic Quantum Architecture and Measurement, Ministry of Education, School of Physics, Beijing Institute of Technology, Beijing 100081, China}
\affiliation{Micronano Center, Beijing Key Lab of Nanophotonics and Ultrafine Optoelectronic Systems, Beijing Institute of Technology, Beijing 100081, China}

\author{Zhiwei Wang}
\email[Corresponding author: ]{zhiweiwang@bit.edu.cn}
\affiliation{Key Laboratory of Advanced Optoelectronic Quantum Architecture and Measurement, Ministry of Education, School of Physics, Beijing Institute of Technology, Beijing 100081, China}
\affiliation{Micronano Center, Beijing Key Lab of Nanophotonics and Ultrafine Optoelectronic Systems, Beijing Institute of Technology, Beijing 100081, China}

\author{Li Lu}
\affiliation{Beijing National Laboratory for Condensed Matter Physics, Institute of Physics, Chinese Academy of Sciences, Beijing 100190, China}
\affiliation{School of Physical Sciences, University of Chinese Academy of Sciences, Beijing 100049, China}
\affiliation{Songshan Lake Materials Laboratory, Dongguan, Guangdong 523808, China}

\author{Fan Yang}
\email[Corresponding author: ]{fanyangphys@tju.edu.cn}
\affiliation{Center for Joint Quantum Studies and Department of Physics, School of Science, Tianjin University, Tianjin 300350, China}
\affiliation{Tianjin Key Laboratory of Low Dimensional Materials Physics and Preparing Technology, Department of Physics, Tianjin University, Tianjin 300350, China}

\date{\today}

\begin{abstract}
TiSe$_2$ is a typical transition-metal dichalcogenide known for its charge-density wave order. In this study, we report the observation of an unusual anisotropic negative magnetoresistance in exfoliated TiSe$_2$ nanoflakes at low temperatures. Unlike the negative magnetoresistance reported in most other transition-metal dichalcogenides, our results cannot be explained by either the conventional two-dimensional weak localization effect or the Kondo effect. A comprehensive analysis of the data suggests that the observed anisotropic negative magnetoresistance in TiSe$_2$ flakes is most likely caused by the three-dimensional weak localization effect.
\end{abstract}

\maketitle

\section{Introduction}\label{sec:Introduction}

The transition metal dichalcogenide (TMD) TiSe$_2$ has attracted considerable research attention over the past decades due to its fascinating feature of charge-density-wave (CDW) order \cite{TiSe2-CDW1, TiSe2-CDW2}. Similar to other layered TMD compounds, the TiSe$_2$ crystal composes of triangular Se-Ti-Se van der Waals layers stacked along the $c$ axis. The CDW transition of TiSe$_2$ occurs at $T_{\textrm{CDW}}\approx200$ K, below which a commensurate $2\times2\times2$ superlattice emerges \cite{TiSe2-CDW1, TiSe2-CDW2}. Although the exact origin of the CDW transition remains controversial, there has been a consensus that it is not due to Fermi surface nesting \cite{TiSe2-CDW-origin1, TiSe2-CDW-origin2, TiSe2-CDW-origin3}. Furthermore, a long-standing debate persists regarding whether TiSe$_2$ is a semimetal \cite{TiSe2-Semimetal-1, TiSe2-Semimetal-2, TiSe2-Semimetal-3} or a narrow-gap semiconductor \cite{TiSe2-Semiconductor-4, TiSe2-Semiconductor-5, TiSe2-Semiconductor-1, TiSe2-Semiconductor-2, TiSe2-Semiconductor-3}. Recent results from angle-resolved photoemission spectroscopy (ARPES) and transport measurements indicate that the CDW state of TiSe$_2$ is a semiconductor with a narrow band gap \cite{TiSe2-Semiconductor-5, TiSe2-Semiconductor-1, TiSe2-Semiconductor-2, TiSe2-Semiconductor-3}. Meanwhile, whether TiSe$_2$ has a finite normal-state energy gap is still under discussion \cite{TiSe2-Semimetal-2, TiSe2-Semimetal-3, TiSe2-Semiconductor-4, TiSe2-Semiconductor-5}.

Compared to studies on the CDW order in TiSe$_2$, experiments focusing on the coherent transport processes have been relatively fewer. Recently, an out-of-plane negative magnetoresistance (NMR) was reported at low temperatures in both bulk samples \cite{TiSe2-Semiconductor-2} and exfoliated flakes \cite{TiSe2-WL} of TiSe$_2$. In the latter case, the observed NMR was attributed to the two-dimensional weak-localization (2D WL) effect \cite{TiSe2-WL}. However, due to the lack of magnetoresistance (MR) data measured at sufficiently low temperatures and under in-plane magnetic fields, further investigations are still needed to fully elucidate the origin of the NMR observed in TiSe$_2$.

In this article, we present our study of the magnetotransport properties of exfoliated TiSe$_2$ nanoflakes. The MR data of TiSe$_2$ flakes were measured down to $T=260$ mK in a $^3$He cryostat. Surprisingly, an unusual anisotropic NMR was observed in \textit{both} perpendicular and in-plane magnetic fields. It was found that the NMR curves measured under in-plane fields were significantly broader than those measured under perpendicular fields, indicating a strong anisotropy of the NMR. Such an unusual anisotropic NMR deviates substantially from the theoretical expectations of both the 2D WL and Kondo effects, which are commonly associated with NMR at low temperatures. A detailed analysis of the data suggests that the observed anisotropic NMR in TiSe$_2$ nanoflakes presumably originates from the 3D WL effect.

\begin{figure}
\includegraphics[width=1 \linewidth]{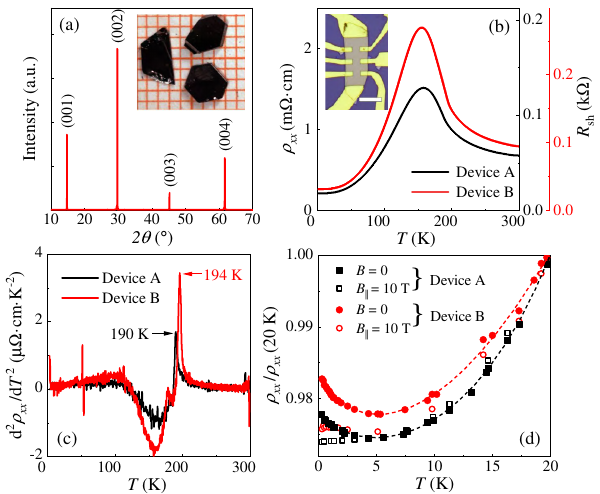}
\caption{\label{fig:Fig1} {
(a) X-ray diffraction pattern of a typical TiSe$_{2}$ crystal, showing Bragg peaks of (00$l$) surfaces. Inset: as-grown TiSe$_2$ crystals. The spacing of grid lines is 1 mm. (b) Temperature dependence of resistivity ($\rho$$_{xx}$) of devices A and B. The values of sheet resistance $R_{\textrm{sh}}$ are shown on the right axes. Inset: optical photo of device A with a 10-$\mu$m scale bar. (c) Second derivative of $\rho_{xx}(T)$ curves. The charge-density-wave transition temperatures were determined to be $T_\textrm{CDW}=190$ K and 194 K for device A and B, respectively. (d) Low-temperature $\rho_{xx}(T)$ curves of devices A and B, normalized to the $\rho_{xx}$ values at $T=20$ K. The resistivity upturn at low temperature is suppressed by an in-plane magnetic field $B_{\parallel}=10$ T.
}}
\end{figure}

\section{Experimental Details}\label{sec:Exp}

\subsection{Crystal Growth}\label{sec:crystalgrowth}

The TiSe$_2$ single crystals used in this work were grown by chemical vapor transport. High-purity titanium powder (99.9\%, Alfa Aesar) and selenium shots (99.99\%, Alfa Aesar) were used as source materials. Before the growth, the source materials were pre-treated to remove any potential oxide layers on the surfaces. The pretreatment includes two steps: hydrogen reduction and sublimation recrystallization. Afterward, the refined titanium and selenium were sealed in a 10-cm-long evacuated quartz tube, with an atomic ratio Ti:Se=1:2.1. During the growth, the temperatures at the cold and hot ends of the quartz tube were optimized to 570 $^{o}$C and 640 $^{o}$C, respectively. The growth took approximately one week. After the growth, TiSe$_2$ crystals with shiny facets were obtained, as shown in the inset of Fig. 1(a). The structure of the obtained TiSe$_2$ crystals was confirmed by X-ray diffraction (XRD). Fig. 1(a) plots the XRD pattern of a typical TiSe$_2$ crystal, in which sharp Bragg peaks of the (00$l$) surfaces are clearly resolved, indicating a high crystallinity of the sample.

\subsection{Device Fabrication and Transport Measurements}\label{sec:devicefab}

Thin TiSe$_2$ flakes were exfoliated onto Si(0.5 mm)/SiO$_2$(300 nm) substrates using Scotch tape. To prevent oxidation, the exfoliation was performed in a glove box filled with N$_2$. After the exfoliation, flakes with regular shapes and flat surfaces were selected out for device fabrication. The Ti (5 nm)/Au (200 nm) electrodes were fabricated onto TiSe$_2$ flakes using electron-beam lithography and electron-beam evaporation. To reduce contact resistance, the contact areas were cleaned \textit{in situ} with Ar plasma prior to metal deposition. The data presented in this paper were obtained in two typical devices, labeled as devices A and B. The thickness of the TiSe$_2$ flakes in devices A and B was measured to be 119 nm and 90 nm, respectively, using an atom force microscope (AFM).

The longitudinal resistivity and Hall resistivity of TiSe$_2$ flakes were measured in a Hall-bar geometry using lock-in amplifiers, with an AC excitation current of 3 $\upmu$A at 17.78 Hz. No obvious overheating caused by excitation current was observed \cite{SM}. The measurements were performed in a $^{3}$He cryostat with a base temperature of $\sim$260 mK.

\begin{figure}
\includegraphics[width=1 \linewidth]{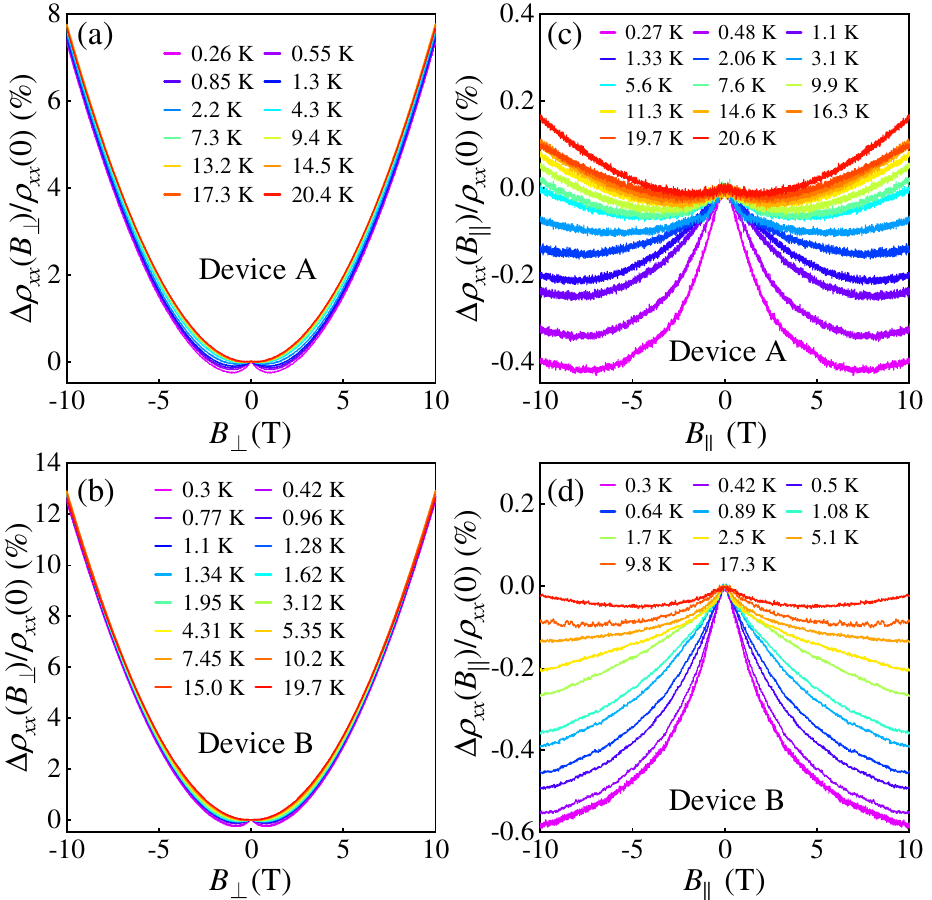}
\caption{\label{fig:Fig2} {
[(a)-(b)] Normalized out-of-plane magnetoresistance of devices A and B, measured at various temperatures. [(c)-(d)] Normalized in-plane magnetoresistance of devices A and B, measured at various temperatures.
}}
\end{figure}

\begin{figure*}
\includegraphics[width=0.95 \linewidth]{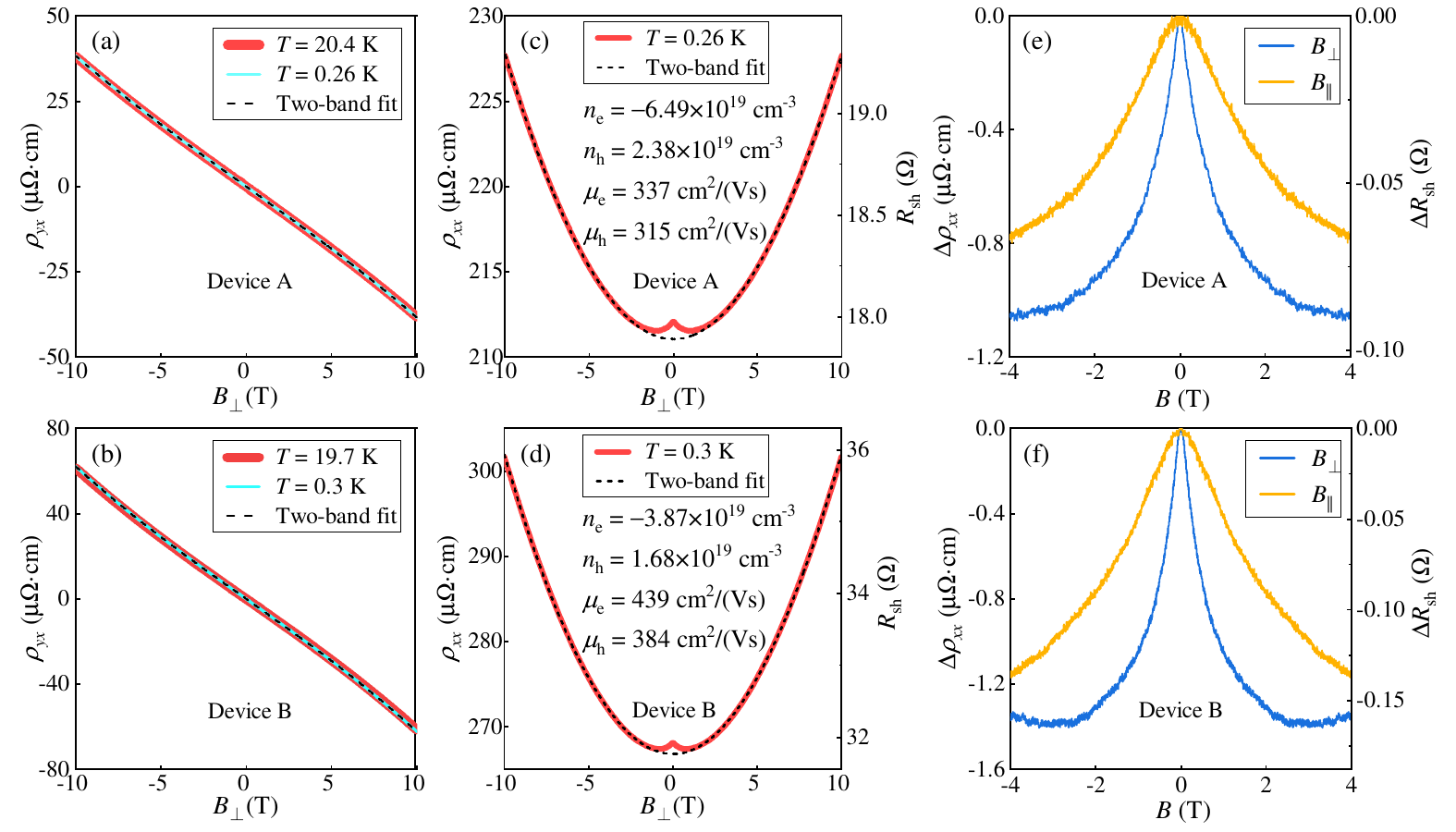}
\caption{\label{fig:Fig3} {
[(a)-(d)] The $\rho_{yx}(B_\perp)$ and $\rho_{xx}(B_\perp)$ curves of device A and B. The values of sheet resistance $R_{\textrm{sh}}$ are shown on the right axes of (c) and (d). Dashed lines are fitted curves obtained using the two-band Drude model. [(e)-(f)] Comparison of the negative magnetoresistivity measured in perpendicular and in-plane magnetic fields. The classical contribution to the out-of-plane magnetoresistivity has been removed by subtracting the fitted curves shown in (c) and (d).
}}
\end{figure*}

\section{Results and discussion}\label{sec:Transport}

\subsection{Temperature Dependence of Resistivity}\label{sec:RT}

The $\rho_{xx}(T)$ curves of TiSe$_2$ flakes behave similarly to those of bulk TiSe$_2$ crystals \cite{TiSe2-CDW1}. As plotted in Fig. 1(b), upon cooling down from room temperature, $\rho_{xx}$ first increases until reaches a maximum at $T\approx155$ K. It then decreases with further lowering of temperature, exhibiting a metallic behavior. The most prominent feature of the $\rho_{xx}(T)$ curves is the anomalous resistivity peak centered around $T\approx155$ K, which has long been a focal point in previous studies. Early work speculated that the resistivity peak might be related to the excitons in the CDW ordering \cite{TiSe2-exciton1, TiSe2-exciton2, TiSe2-exciton3}. Recently, Watson \textit{et al.} proposed that it could be explained by a simple two-band model \cite{TiSe2-twoband}, without invoking the CDW order in TiSe$_2$.

Similar to bulk crystals, TiSe$_2$ flakes also undergo a CDW transition during cooling. As shown in Fig. 1(c), the transition temperatures for devices A and B were determined to be $T_\textrm{CDW}=$ 190 K and 194 K, respectively, based on the peaks in the second derivative of the $\rho_{xx}(T)$ curves. These values are in good agreement with the $T_\textrm{CDW}$ previously reported in bulk crystals \cite{TiSe2-CDW1} and exfoliated flakes \cite{TiSe2-WL} of TiSe$_2$ .

Fig. 1(d) shows the low-temperature part of the $\rho_{xx}(T)$ curves normalized to the value at $T=20$ K. At zero magnetic field, the resistivity of TiSe$_2$ flakes does not saturate at low temperatures but instead increases slightly with decreasing $T$, leading to a resistivity upturn at $T<5$ K. The low-temperature resistivity upturns are completely suppressed by an in-plane field $B_\parallel=10$ T.

Possible mechanisms responsible for such low-temperature resistivity upturns include WL, the Kondo effect, and electron-electron interactions (EEI). Among them, WL and the Kondo effect are sensitive to magnetic fields, whereas EEI is not \cite{EEI}. Therefore, the suppression of the resistivity upturns by external fields rules out EEI as their cause.

\subsection{Magnetotransport}\label{sec:MR}

The MR curves of devices A and B were measured in both perpendicular and in-plane magnetic fields at temperatures ranging from 0.26 K to 20.6 K, as plotted in Fig. 2. All curves exhibit distinct NMR in the low-field region. The observed low-field NMR is highly temperature-dependent. It flattens rapidly as $T$ increases and almost disappears at $T\approx20$ K. Such temperature-sensitive NMR is typically attributed to either the Kondo effects \cite{Kondo-1, Kondo-2} or WL \cite{TiSe2-WL}. The origin of the NMR will be discussed in detail in the following sections.

At high fields, the out-of-plane MR is dominated by the classical parabolic MR, which shows little variation between $T=$0.3 K and 20 K, as presented in Figs. 2(a) and (b). Since the amplitude of classical MR is primarily determined by the concentration and mobility of charge carriers, such a behavior indicates that the carrier concentration and mobility of TiSe$_2$ flakes do not change significantly over this temperature range.

At $T\leq20$ K, the Hall resistivity [$\rho_{yx}(B)$] is almost independent of temperature, as illustrated in Figs. 3(a) and (b), further confirming that the carrier density of TiSe$_2$ remains constant at low temperatures. The slightly curved $\rho_{yx}(B)$ curves suggest the presence of multiple types of charge carriers in TiSe$_2$, consistent with the results from ARPES measurements~\cite{TiSe2-Semiconductor-5}.

\subsection{Possible Origins of Negative Magnetoresistance}\label{sec:Discussion}

\begin{figure}
\includegraphics[width=1 \linewidth]{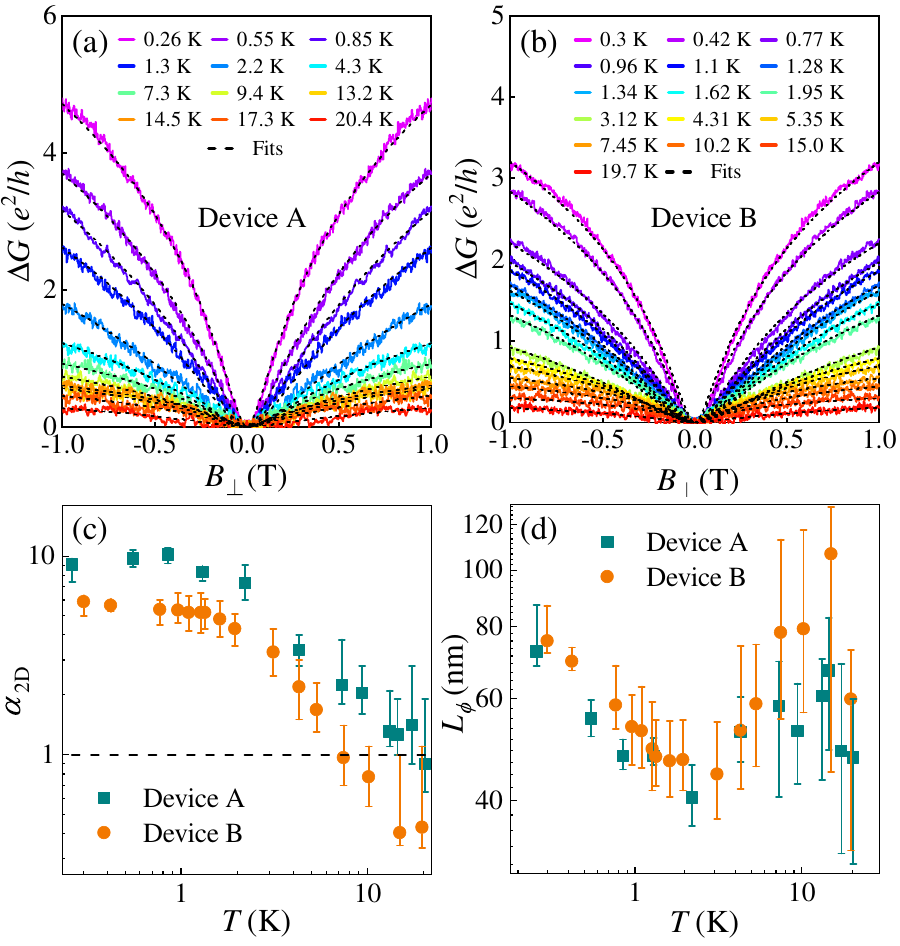}
\caption{\label{fig:Fig4} {
[(a)-(b)] Low-field magnetoconductance of devices A and B at various temperatures. Here $\Delta G(B)$ is the change of sheet conductance after subtracting the classical background. Dashed lines represent the best fits to the data using the HLN formula [Eq. (3)]. [(c)-(d)] The obtained fitting parameters $\alpha_\textrm{2D}$ and $L_{\phi}$ plotted as a function of temperature. Error bars represent the ranges within which the fitting results appear acceptable by eye. At low temperatures, $\alpha_\textrm{2D}$ significantly exceeds the threshold $\alpha_\textrm{2D}=1$ expected from the 2D weak localization theory.}}
\end{figure}

\begin{figure*}
\includegraphics[width=0.9 \linewidth]{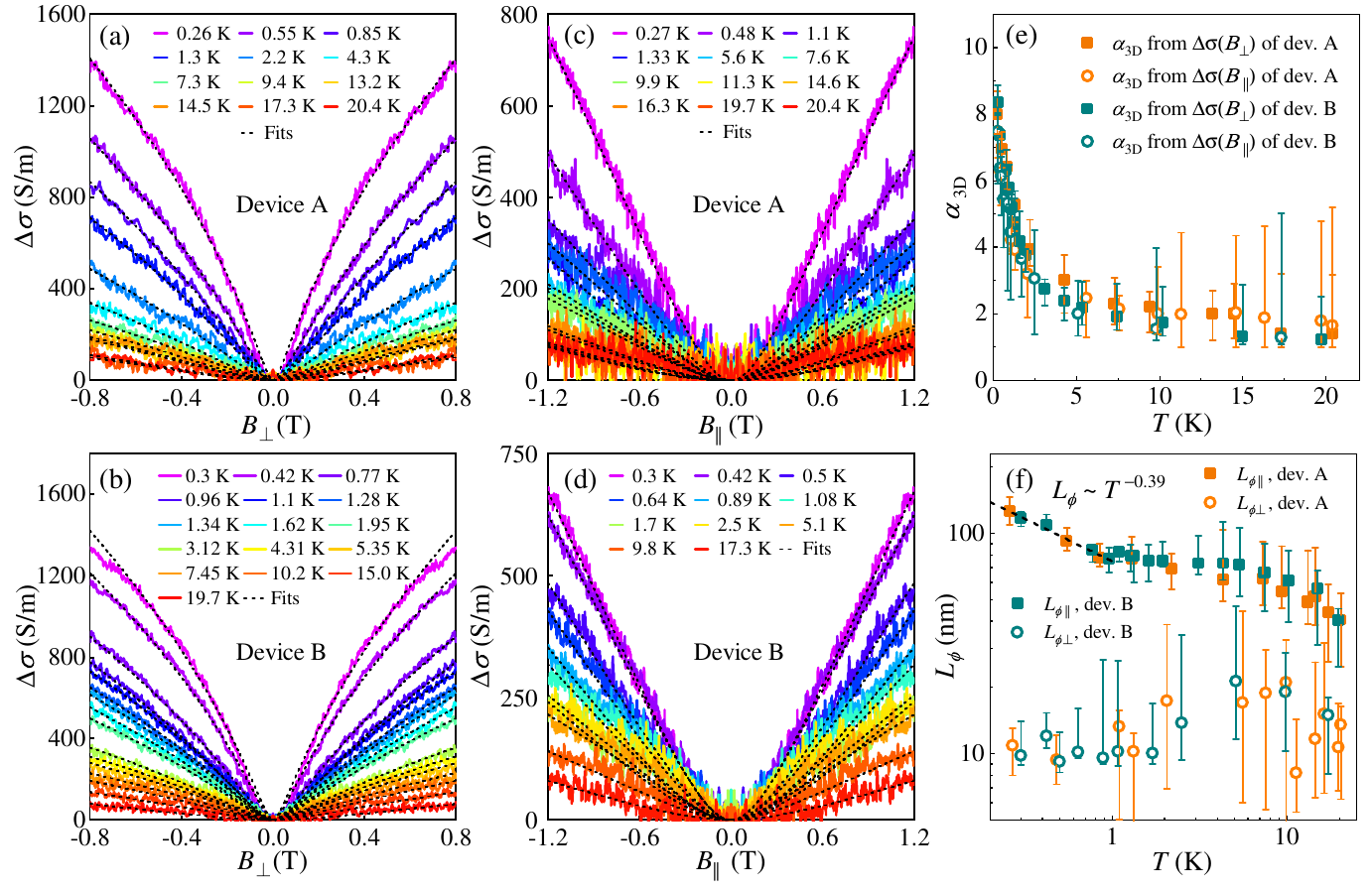}
\caption{\label{fig:Fig5} {
Fitting to the magnetoconductivity data using the 3D weak localization theory. [(a)-(d)] Low-field $\Delta \sigma(B)$ curves of devices A and B, measured in perpendicular and in-plane magnetic fields and at various temperatures. Here $\Delta \sigma \equiv \sigma(B) - \sigma(0)$. Dashed lines are fits to the $\Delta\sigma(B)$ data using the 3D weak localization theory [Eq. (5)]. [(e)-(f)] Temperature dependence of obtained fitting parameters $\alpha_{\textrm{3D}}$, $L_{\phi\parallel}$ and $L_{\phi \perp}$ for devices A and B. Error bars indicate the ranges within which the deviation between the fitting curves and the data looks acceptable.
}}
\end{figure*}

\subsubsection{The Kondo Effect}\label{sec:Kondo}

The NMR caused by the Kondo effect is expected to be isotropic in magnetic fields. To reliably compare the NMR data taken under perpendicular and in-plane magnetic fields, we first subtract the background of classical MR in the $\rho_{xx}(B_{\perp})$ curves. The classical MR is obtained by fitting to the $\rho_{xx}(B_{\perp})$ and $\rho_{yx}(B_{\perp})$ data simultaneously using a two-band Drude model \cite{twoband-Drude1, twoband-Drude2} consisting of an electron band and a hole band:
\begin{equation}
\rho_{xx}=\frac{1}{e}\left[\frac{|n_{\rm h}| \mu_{\rm h}+|n_{\rm e}| \mu_{\rm e}+B^2 \mu_{\rm h} \mu_{\rm e} (|n_{\rm h}| \mu_{\rm e}+|n_{\rm e}| \mu_{\rm h} )}{(|n_{\rm h}| \mu_{\rm h}+|n_{\rm e}| \mu_{\rm e} )^2+B^2 \mu_{\rm h}^2 \mu_{\rm e}^2 (n_{\rm h}+n_{\rm e} )^2 }\right]
\end{equation}
and
\begin{equation}
\rho_{yx}=\frac{B}{e}\left[\frac{n_{\rm h} \mu_{\rm h}^2+n_{\rm e}\mu_e^2+B^2 \mu_{\rm h}^2 \mu_e^2 (n_{\rm h}+n_{\rm e})}{(|n_{\rm h}| \mu_{\rm h}+|n_{\rm e}| \mu_{\rm e} )^2+B^2 \mu_{\rm h}^2 \mu_e^2 (n_{\rm h}+n_{\rm e})^2 }\right],
\end{equation}
where $e$ is the elementary charge, $n_{\rm e}$, $n_{\rm h}$, $\mu_{\rm e}$, and $\mu_{\rm h}$ correspond to the 2D carrier density and mobility of the electron and hole bands, respectively. Here, the electron band represents the 3D-like electron pocket at the $L$ point, while the hole band is associated with the 2D-like hole pocket at the $\Gamma$ point \cite{twoband-Drude1}. The fitting results are plotted as dashed lines in Figs. 3(a)-(d), and more details are available in the Supplemental Material \cite{SM}.

After subtracting the background, the net resistivity change $\Delta\rho_{xx}$ due to the out-of-plane NMR is obtained and plotted together with the $\Delta\rho_{xx}(B_{\parallel})$ data in Figs. 3(e) and (f). The procedure for the classical-background subtraction is presented in the Supplemental Material \cite{SM}. Clearly, for both devices, the $\Delta\rho_{xx}(B_{\parallel})$ curves are much broader than the $\Delta\rho_{xx}(B_{\perp})$ curves, indicating a high degree of anisotropy in the NMR. Such anisotropy is inconsistent with the expectations of the Kondo effect, which predicts an isotropic NMR independent of field direction \cite{Kondo-1, Kondo-2}. Therefore, the possibility of the Kondo effect can be safely excluded.

\subsubsection{Two-Dimensional Weak Localization}\label{sec:2DWL}

In previous work, the NMR in TiSe$_2$ flakes was attributed to the 2D WL effect \cite{TiSe2-WL}. However, in our experiment, the NMR is observed under both perpendicular and in-plane magnetic fields, as shown in Fig. 2. The presence of in-plane NMR strongly points to a 3D origin, rendering the explanation of 2D WL unlikely.

To further investigate whether the observed NMR can be ascribed to the 2D WL effect, we carefully analyzed the magnetoconductance (MC) data using the 2D WL theory. In this theory, the low-field MC due to 2D WL is described by the simplified Hikami-Larkin-Nagaoka (HLN) equation \cite{HLN}
\begin{equation}
\Delta G=\alpha_\textrm{2D}\frac{e^2}{2 \pi^2 \hbar} \left[ \psi\left(\frac{1}{2} + \frac{\hbar}{4eL_{\phi}^2B}\right) - \ln\left(\frac{\hbar}{4eL_{\phi}^2B}\right) \right],
\end{equation}
where $\Delta G(B)$ is the change in sheet conductance under magnetic fields, $L_\phi$ is the phase coherence length, $\psi$ is the digamma function, and $\alpha_\textrm{2D}$ is a prefactor determined by the number of independent diffusive transport channels.

To check whether the observed NMR can be attributed to 2D WL, we fitted the HLN equation to the low-field MC data, as plotted in Fig 4. The contribution from classical MR had been removed prior to the fitting using the method described in Sec. \ref{sec:Kondo}. Although the fitting results look reasonably good, the extracted parameters do not align with the theoretical expectations. First, the parameter $\alpha_\textrm{2D}$ was found to increase monotonically with decreasing $T$, reaching $\alpha_\textrm{2D}\approx 10$ in device A at $T=0.26$ K, far beyond the expected value of $\alpha_\textrm{2D}=1$ for 2D WL in a multi-band system with inter-band scattering. In addition, at $T>3$ K, the $L_\phi(T)$ data are highly scattered and lack a clear trend, which is inconsistent with the power-law temperature dependence predicted by the 2D WL theory.

We also note that similar large values of $\alpha_{\mathrm{2D}}$, which far exceeds the expected value for 2D weak antilocalization (WAL), have been reported in several bulk topological semimetals \cite{WL-in-TSM-1, WL-in-TSM-2, WL-in-TSM-3}. As in our case, these observations cannot be explained by conventional 2D WAL and instead point to a 3D origin.

Based on the analysis presented above, we conclude that the NMR observed in this experiment does not originate from the 2D WL effect.

\subsubsection{Anisotropic Three-Dimensional Weak Localization}\label{sec:3DWL}

After ruling out the possibility of the Kondo and 2D WL effects, the only remaining known mechanism to explain the observed NMR in TiSe$_2$ flakes is the 3D WL effect. Unlike the 2D WL case, the theory of 3D WL predicts the occurrence of NMR in both out-of-plane and in-plane magnetic fields. For anisotropic systems with $D_{\perp}\neq D_{\parallel}$, where $D_{\perp}$ and $D_{\parallel}$ are the diffusion coefficients in the out-of-plane and in-plane directions, the low-field conductivity correction due to 3D WL is given by \cite{3D-WL-1, 3D-WL-2}
\begin{equation}\label{eqn:eq4}
\Delta \sigma(B_{\perp}) = \alpha_{\textrm{3D}}\frac{e^2}{2 \pi^2 \hbar} \sqrt{\frac{eB_{\perp}}{\hbar}} F\left(\frac{\hbar}{4eL_{\phi\parallel}^2 B_{\perp}}\right)
\end{equation}
and
\begin{equation}\label{eqn:eq5}
\Delta \sigma(B_{\parallel}) = \alpha_{\textrm{3D}}\frac{e^2}{2 \pi^2 \hbar} \sqrt{\frac{eB_{\parallel}}{\hbar}} F\left(\frac{\hbar}{4eL_{\phi\parallel} L_{\phi\perp} B_{\parallel}}\right).
\end{equation}
Here, $L_{\phi\perp}$ and $L_{\phi\parallel}$ denote the phase coherence lengths in the perpendicular and in-plane directions, $\alpha_{\textrm{3D}}=\sqrt{D_{\parallel}/D_{\perp}}$ is the anisotropy parameter, and the function $F$ is expressed as
\begin{align*}
F(x) &= \sum_{n=0}^{\infty} \left\{ 2 \left[ (n+1+x)^{0.5} - (n+x)^{0.5} \right] \right. \\
&\quad \left. - (n+0.5+x)^{-0.5} \right\}.
\end{align*}

Figs. 5(a)-(d) show the 3D magnetoconductivity data $\Delta\sigma(B)$ for both devices. As usual, the background of classical MR has been removed. Notably, the amplitudes of the $\Delta\sigma(B)$ curves are very similar in the two devices, despite the distinct differences in the amplitudes of the $\Delta G(B)$ curves shown in Figs. 4(a)-(b). This strongly implies that the observed NMR is essentially due to a quantum correction to the conductivity, rather than to the conductance, which further confirms the 3D nature of the NMR.

To determine whether the observed NMR can be well described by the 3D WL thoery, we analyzed the low-field $\Delta\sigma(B)$ data from devices A and B using Eqs. (4) and (5). Here, it is important to note that the anisotropy parameter $\alpha_{\textrm{3D}}$ appears in both Eqs. (4) and (5) and therefore should not be taken as an independent parameter when fitting to the $\Delta\sigma(B_{\perp})$ and $\Delta\sigma(B_{\parallel})$ curves. However, in this experiment, the $\Delta\sigma(B_{\perp})$ and $\Delta\sigma(B_{\parallel})$ curves were not measured at exactly the same temperature. Therefore, during the fitting, we aimed to collapse the $\alpha_{\textrm{3D}}$ data extracted from the $\Delta\sigma(B_{\perp})$ and $\Delta\sigma(B_{\parallel})$ curves onto a common $\alpha_{\textrm{3D}}(T)$ trend line to ensure consistency.

The 3D WL analysis yields consistent results. For both devices, the fitting curves agree well with the low-field magnetoconductivity data, as shown in Figs. 5(a)-(d). However, surprisingly, unlike the 3D WL effect previously reported in other materials \cite{3D-WL-3, 3D-WL-Diamond}, the anisotropy parameter $\alpha_{\textrm{3D}}$ obtained in this experiment is not constant, but highly temperature-dependent. As shown in Fig. 5(e), at low temperatures, the extracted $\alpha_{\textrm{3D}}$ increases sharply from $\alpha_{\textrm{3D}}\approx2$ at $T=5$ K to $\alpha_{\textrm{3D}}\approx8$ at $T=0.26$ K.

Clearly, such a highly temperature-sensitive anisotropy parameter cannot be attributed to the spatial anisotropy of the effective mass tensor. Although the exact origin of such an anomalous temperature dependence of $\alpha_{\textrm{3D}}$ is not yet clear, we speculate that it may be arise from a possible temperature-dependent anisotropy in the inelastic scattering rate in TiSe$_2$.

Fig. 5(f) shows the $L_{\phi}(T)$ data extracted from the fitting. As $T$ decreases, $L_{\phi\parallel}$ increases monotonically, whereas $L_{\phi\perp}$ shows no clear trend. At low temperatures, $L_{\phi\perp}$ falls in the range of 10–20 nm, which is significantly smaller than the flake thickness of approximately 100 nm. This indicates that, in terms of diffusive transport, the TiSe$_2$ flakes should be regarded as 3D systems, consistent with the presence of 3D WL.

The extracted values of $L_{\phi\parallel}$ are comparable to those typically obtained in other materials, whereas the values of $L_{\phi\perp}$ are noticeably smaller. Such a pronounced anisotropy in phase coherence length may originate from both the anisotropic effective mass and scattering rates of electrons in TiSe$_2$, although the exact mechanism remains to be clarified in future studies. If our interpretation is correct, a dimensional crossover from 3D to 2D WL should occur when the flake thickness is reduced below $\sim$20 nm. We plan to investigate this transition in future work.

The in-plane phase coherence length $L_{\phi\parallel}$ reaches about 100 nm at low temperatures and follows a temperature dependence of $L_{\phi\parallel} \sim T^{-0.39}$ below 1 K, as shown by the dashed line in Fig. 5(f). Theoretically, the exponent $p$ in the relation $L_{\phi\parallel}\sim T^{-p}$ is expected to take values of $p=1.5$, 1 and 0.75 for systems in which the inelastic scattering is dominated by electron-phonon interaction, EEI in the clean limit and EEI in the dirty limit, respectively \cite{EEI}. Clearly, the exponent $p=0.39$ obtained in this experiment does not match any of these theoretically predicted values. Although a similar value of $p=0.35$ has been reported in the 3D WL of nitrogen-doped ultrananocrystalline diamond films \cite{3D-WL-Diamond}, the exact origin of this deviation remains unclear. Further research is needed to fully elucidate the discrepancy.

\section{Conclusion}

In summary, we observed an unusual anisotropic NMR in exfoliated TiSe$_2$ nanoflakes. Unlike the NMR commonly seen in other 2D materials, our results cannot be explained by either the 2D WL effect or the Kondo effect. We tentatively attribute the observed NMR to the anisotropic 3D WL effect.
Our results provide direct experimental signatures of phase-coherent transport processes in TiSe$_2$ and may be helpful for advancing the understanding of phase-coherent transport phenomena in other layered TMD materials with similar anisotropic characteristics.

\begin{acknowledgments}

This work is supported by the National Key Research and Development Program of China (Grant Nos. 2022YFA1403400, 2020YFA0308800); by the Beijing Natural Science Foundation (Grant No. Z210006); by the Beijing National Laboratory for Condensed Matter Physics (Grant No. 2023BNLCMPKF007).

\end{acknowledgments}

\clearpage
\onecolumngrid
\appendix
\renewcommand{\thesection}{S\arabic{section}}
\renewcommand{\thefigure}{S\arabic{figure}}
\renewcommand{\thetable}{S\arabic{table}}
\renewcommand{\theequation}{S\arabic{equation}}
\setcounter{figure}{0}
\setcounter{table}{0}
\setcounter{equation}{0}
\setcounter{section}{0}

\begin{center}
{\large\bfseries
\setlength{\baselineskip}{2\baselineskip}%
Supplemental Material
}
\end{center}

\section{Comparison of Magnetoresistance Data under Different Currents}

To rule out the possibility of overheating caused by the excitation current, we compared the magnetoresistance data measured at base temperature using three different excitation currents: 1 $\upmu$A, 3 $\upmu$A and 5 $\upmu$A. As shown in Fig. S1(a), the magnetoresistance curves for 1 $\upmu$A and 3 $\upmu$A are closely matched, whereas the curve for 5 $\upmu$A appears noticeably flatter. These results indicate that a 3 $\upmu$A excitation current does not cause overheating in our TiSe$_2$ samples. Additionally, the noise level at 3 $\upmu$A is significantly lower than that at 1 $\upmu$A. Therefore, we select a 3 $\upmu$A excitation current for our main measurements.

\section{Estimation of Uncertainty in Resistivity Calculation}

The length-to-width ratio of the devices was calculated using $l_1/w$, where $l_1$ is the distance between the midpoints of the two voltage (potential) contacts, and $w$ is the width of the TiSe$_2$ flake, as illustrated in Figs. S1(b) and (c). The selected TiSe$_2$ flakes have well-defined geometries, which minimize the uncertainty in estimating the length-to-width ratio due to sample shape. Although the presence of voltage contacts may perturb the current distribution and introduce some uncertainty in determining the effective length, this uncertainty can be reasonably estimated by taking $l_2$ and $l_3$ as the upper and lower bounds for the length, respectively, as indicated in Figs. S1(b) and (c). For both devices, the maximum relative uncertainty in the length-to-width ratio is approximately $\pm20\%$, which would correspondingly result in an $\sim20\%$ uncertainty in the calculated resistivity $\rho_{xx}$, as well as in the prefactor $\alpha$. However, this level of uncertainty does not affect the validity of our main conclusions.

\section{Fitting Details of the Classical Magnetoresistance Background}

To reliably obatin the classical parabolic background in the $\rho_{xx}(B_{\perp})$ curves, we fitted the two-band Drude model simultaneously to the high-field ($B_{\perp}>3$ T) $\rho_{xx}(B_{\perp})$ and $\rho_{yx}(B_{\perp})$ data. According to the two-band Drude model, the longitudinal and transverse resistivity are respectively given by
\begin{equation}\label{eq1}
  \rho_{xx}  = \left( \frac{1}{e} \right) \frac{|n_1| \mu_1 + |n_2| \mu_2 + B_{\perp}^2 \mu_1 \mu_2 (|n_1| \mu_2 + |n_2| \mu_1)}{(|n_1| \mu_1 + |n_2| \mu_2)^2 + B_{\perp}^2 \mu_1^2 \mu_2^2 (n_1 + n_2)^2}
\end{equation}
and
\begin{equation}\label{eq2}
  \rho_{yx}  = \left( \frac{B_{\perp}}{e} \right) \frac{(n_1 \mu_1^2 + n_2 \mu_2^2) + B_{\perp}^2 \mu_1^2 \mu_2^2 (n_1 + n_2)}{(|n_1| \mu_1 + |n_2| \mu_2)^2 + B_{\perp}^2 \mu_1^2 \mu_2^2 (n_1 + n_2)^2},
\end{equation}
where $n_1$, $n_2$, $\mu_1$ and $\mu_2$ are carrier density and mobility of the two bands. Note that here $n_1$, $n_2$ are sign-carrying variables, where a positive value indicates p-type carriers and a negative value indicates n-type carriers.

\begin{table}[b]
\caption{\label{tab:s1}Fitted parameters of the two-band Drude model.}
\begin{ruledtabular}
\begin{tabular}{lcccc}
 & $n_1$ ($10^{19}$ cm$^{3}$) & $n_2$ ($10^{19}$ cm$^{3}$) & $\mu_1$ (cm$^{2}$V$^{-1}$s$^{-1}$) & $\mu_2$ (cm$^{2}$V$^{-1}$s$^{-1}$) \\
\hline
Fitted values   & $-6.49$ & $2.38$ & $337$ & $315$ \\
Upper bounds    & $-6.20$ & $2.73$ & $345$ & $340$ \\
Lower bounds    & $-6.79$ & $2.07$ & $329$ & $292$ \\
\end{tabular}
\end{ruledtabular}
\end{table}

Figure S2 illustrates the two-band fitting of the high-field $\rho_{xx}(B_{\perp})$ and $\rho_{yx}(B_{\perp})$ curves measured at base temperature. To estimate the uncertainty range of the fitting parameters, we vary $n_1$ around its optimal value and then optimize the other parameters to minimize the residuals (defined as $\rho-\rho^{\textrm{fit}}$) between the measured and fitted curves. We then compare the standard deviation of the residuals with its minimized value obtained under the best-fit condition. The boundary of the parameter range is defined as the point at which the standard deviation of the residuals reaches ten times its best-fit value. The fitted values and uncertainty ranges of all fitting parameters are listed in Table S1.

\section{Procedure for Classical Magnetoresistance Background Subtraction}

As described in the main text, the raw $\rho_{xx}(B_{\perp})$ data contain a large positive classical magnetoresistance (MR) in addition to a smaller quantum correction. To isolate the quantum contribution, we fitted the two-band Drude model to the high-field part of the $\rho_{xx}(B_{\perp})$ data, thereby obtaining the corresponding classical background $\rho_{\mathrm{fit}}(B_{\perp})$. The preliminary background-subtracted curve is then defined as
\begin{equation}
    \Delta\rho_{xx}^{\prime}(B_{\perp}) = \rho_{xx}(B_{\perp}) - \rho_{\mathrm{fit}}(B_{\perp}).
\end{equation}

Since this subtraction procedure uses the high-field region as the reference point, $\Delta\rho_{xx}^{\prime}(B_{\perp})$ does not naturally vanish at $B=0$. To make the result directly comparable with theoretical descriptions of quantum corrections (such as the Kondo effect), which characterize the field-induced relative change of resistivity from its zero-field value, we further shift the curve by subtracting $\Delta\rho_{xx}^{\prime}(0)$:
\begin{equation}
    \Delta\rho_{xx}(B_{\perp}) = \Delta\rho_{xx}^{\prime}(B_{\perp}) - \Delta\rho_{xx}^{\prime}(0).
\end{equation}

This procedure ensures that $\Delta\rho_{xx}(0)=0$, consistent with the theoretical framework. The resulting $\Delta\rho_{xx}(B_{\perp})$ curves shown in Figs.~3(e)--(f) of the main text therefore represent the relative change of resistivity under magnetic field after removing the classical background. For illustration, Fig.~S3 shows an example of this background subtraction process using the data from Device~A.

\begin{figure}
\includegraphics[width=0.7 \linewidth]{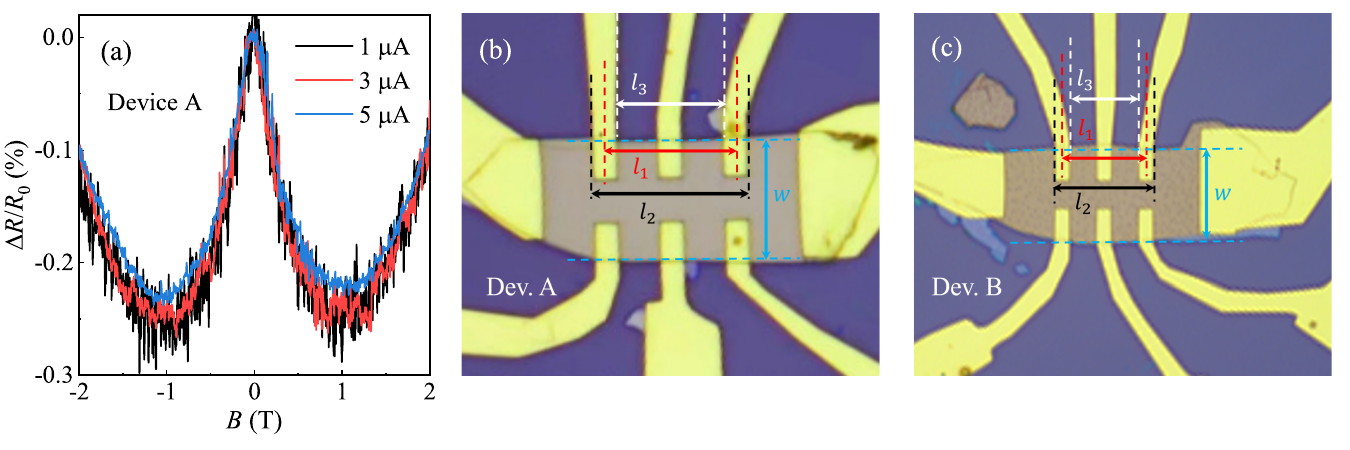}
\caption{\label{fig:FigS1} {
(a) Raw magnetoresistance data of device A measured at base temperature under three different excitation currents. The curves appear slightly different from those shown in Fig. 3(e) in the main text. This is because the classical background has been subtracted in Fig. 3(e), whereas it remains in the raw data presented here. (b)-(c) Schematic illustration of how the uncertainty in the length-to-width ratio of the flakes is estimated. The optical image of device B was taken after it had been kept in air for long time after the measurements, and therefore the surface of the flake got slightly oxidized.
}}
\end{figure}

\begin{figure}
\includegraphics[width=0.95 \linewidth]{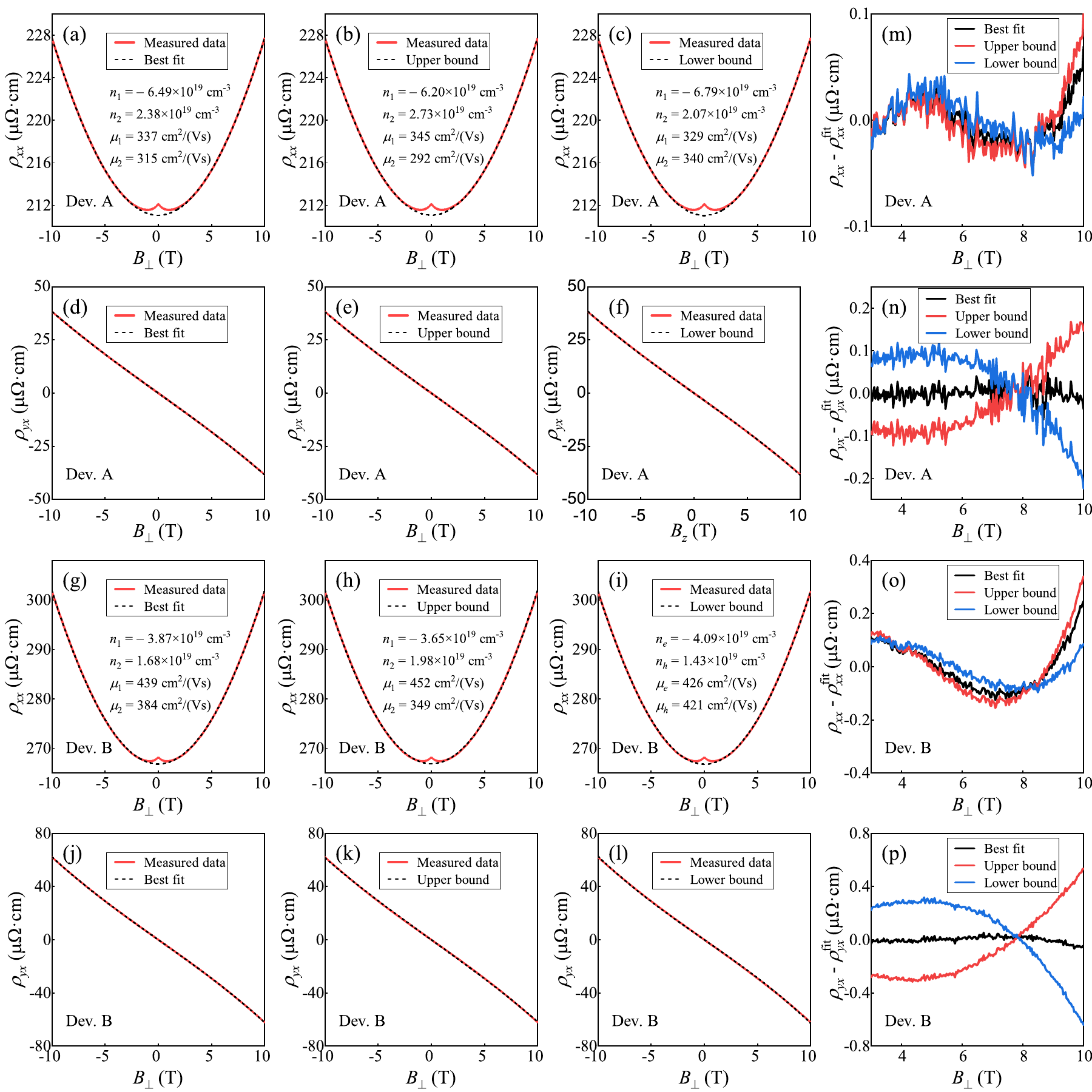}
\caption{\label{fig:FigS2} {
(a)-(l) The measured $\rho_{xx}(B_{\perp})$ and $\rho_{yx}(B_{\perp})$ curves (solid lines) at base temperature, together with the corresponding fitted curves (dashed lines) based on the two-band Drude model. (m)-(p) Residuals between the measured and fitted $\rho_{xx}(B_{\perp})$ and $\rho_{yx}(B_{\perp})$ curves.
}}
\end{figure}

\begin{figure}
\includegraphics[width=0.9 \linewidth]{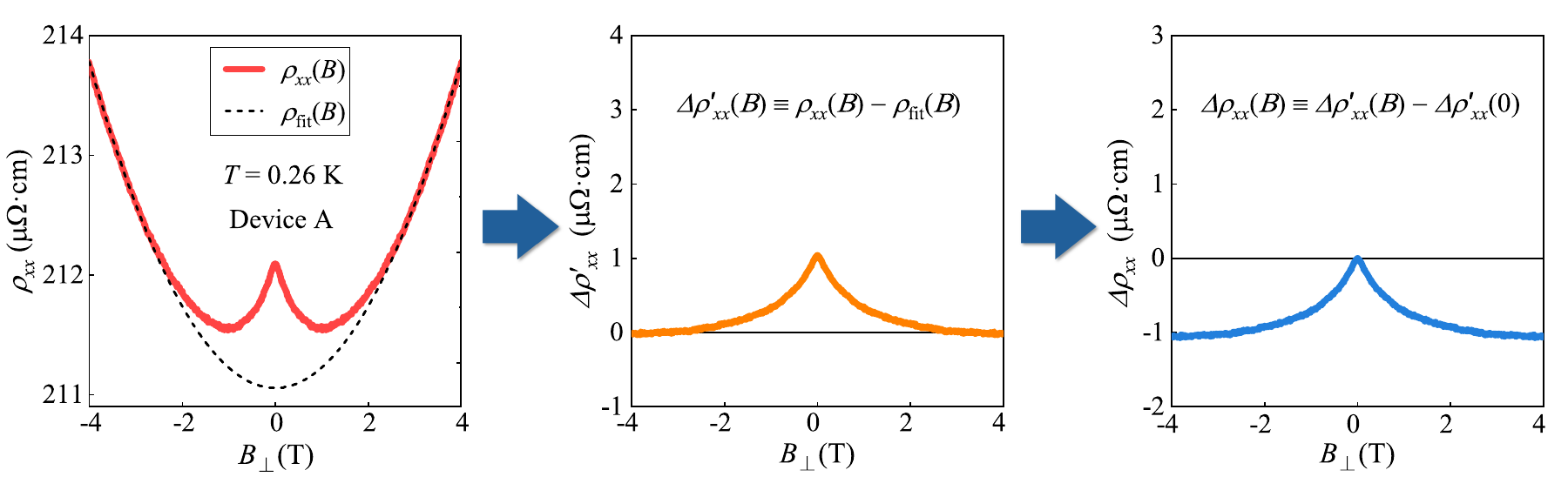}
\caption{\label{fig:FigS3} {
Illustration of the classical magnetoresistance background subtraction procedure, using the data from Device~A as an example.
}}
\end{figure}

\end{document}